\documentclass[aps,prd,12pt,
nofootinbib,tightenlines,superscriptaddress,preprintnumbers]{revtex4}

\usepackage{graphicx}
\usepackage{rotating}
\usepackage{latexsym}

\newcommand{\beq}{\begin{equation}}
\newcommand{\eeq}{\end{equation}}
\newcommand{\bqa}{\begin{eqnarray}}
\newcommand{\eqa}{\end{eqnarray}}

\def\slash{\hskip-0.7em/ \,}
\def\slashh{\hskip-0.5em/ }

\begin{document}

\preprint{BI-TP 2008/05}
\preprint{HD-THEP 2008/14}

\vspace{1.0cm}

\title{Jet quenching and broadening: the  transport coefficient
 $\hat{q}$ \\ in an anisotropic plasma}

\author{Rudolf~Baier\footnote{E-mail address:
    baier@physik.uni-bielefeld.de}}
\affiliation{ Physics Department, University of Bielefeld, D-33501 Bielefeld,
Germany}
\author{
Yacine~Mehtar-Tani\footnote{E-mail address: Y.Mehtar-Tani@thphys.uni-heidelberg.de}}
\affiliation{Institut f\"ur Theoretische Physik, Universit\"at Heidelberg,
 D-69120 Heidelberg, Germany}  


\begin{abstract}

 \centerline{\bf{Abstract}}
The jet quenching  parameter $\hat{q}$ is analyzed for a quark jet
propagating in an anisotropic plasma. The momentum anisotropy is 
calculated at high temperature 
of the underlying quark-gluon plasma. $\hat{q}$ is explicitly 
estimated in leading-logarithmic approximation by
the broadening of the massless quark interacting via gluon exchange.
A plasma instability  is present. Strong indications are found
that $\hat{q}$ is increasing with increasing anisotropy.
Possible implications for the saturation scale $Q_s$
in $A-A$ collisions are pointed out.
\end{abstract}

\maketitle

\section{Introduction}

Jet quenching in $Au-Au$ collisions at RHIC \cite{Collaboration:2008qa,RHIC},
 i.e. the suppression of
$\pi^0$ and $\eta$ production at large transverse momenta in
central collisions when
compared  with scaled measurements in $p-p$ collisions 
\cite{Tannenbaum:2006ch}
is considered as  one of the important observations in favour of
the production of a plasma state \cite{Muller:2007rs,CasalderreySolana:2007pr}.

The experimental data from RHIC are used to determine the actual value
of the transport coefficient $\hat{q}$, which controls the radiative
energy loss, responsible for jet quenching in the
induced  gluon bremsstrahlung picture
\cite{Gyulassy:1993hr,Baier:1996sk,Baier:1996kr,Baier:2000mf,Wiedemann:2000tf,
Zakharov:2004vm,Arnold:2002ja,Turbide:2005fk,Arleo:2002kh,Arleo:2006xb,
Kovner:2003zj,Salgado:2005zp,MehtarTani:2006xq}.

Within the kinetic theory framework the coefficient 
(for an isotropic medium) is estimated by

\beq
\hat{q} = \rho \int d^2q_\perp q^2_\perp \frac{d \sigma}{d^2q_\perp} ~,
\label{eq:basic}
\eeq
where $\rho$ is the number density of the constituents of the medium,
and $\frac{d \sigma}{d^2q_\perp}$ is the differential  scattering cross section of
the parton (massless quark or gluon) on the medium.

\noindent
It is important to keep in mind that $\hat{q}$ 
not only determines the energy loss, but it is also related to
$p_\perp$-broadening (per unit length)
 of energetic partons propagating in the medium/plasma
\cite{Baier:1996sk}.

Because of many theoretical uncertainties in the determination of $\hat{q}$
its value maybe quoted to be in a broad range of $0.5 - 20 ~GeV^2/fm$,
see e.g \cite{Eskola:2004cr, Arleo:2002kh,Baier:2006fr}.

It has been found (see e.g.
\cite{Mrowczynski:1993qm,Randrup:2003cw,Mrowczynski:2005ki,
Romatschke:2003yc,Romatschke:2003ms,Romatschke:2003vc,
Romatschke:2004jh,Romatschke:2004au,Arnold:2004ti,Strickland:2006pb})
that the physics of anisotropic plasmas
 differs from that of isotropic
ones, because of the presence of plasma instabilities in the former.
Due to these observations
it requires to reanalyze 
 $\hat{q}$  in the context of anisotropic
plasmas.
 Momentum broadening in a homogeneous
but locally anisotropic high-temperature system for a heavy quark induced by collisions 
has been discussed recently \cite{Romatschke:2006bb} (see also
\cite{Romatschke:2004au,Moore:2004tg,Dumitru:2007rp}).
Here the case for massless partons is of relevance.

\noindent
In the following the anisotropic medium is modeled as a high temperature 
 gas. As pointed out in \cite{Romatschke:2006bb} the temperature is kept as a
 placeholder of the correct hard scale, like the saturation scale $Q_s$,
 present in the anisotropic system.
 These  assumptions allow an analytic and
tractable treatment, at least for small anisotropies,
considered in this paper.
In detail the main effect is due to the propagation of the 
dominantly soft gluon in this 
medium.

\section{Setup}

In the following momentum broadening (per unit pathlength) for a massless quark
in the high energy limit (eikonal limit) in a homogeneous but anisotropic high
temperature plasma is studied. 
 The following kinematical setup is used:
the initial energetic (hard) quark propagates with momentum $p^\mu$ along the $z$-axis,
whereas the beam nuclei collide along the $y$-axis, which is the
direction of anisotropy, denoted by the three-dimensional unit vector
${\bf{n}} = (0,1,0)$.

In the limit under consideration 
momentum broadening $q_\perp$ takes place only in the $x-y$ plane,
$q_\perp^2 = q_x^2 + q_y^2$. 

Instead of kinetic theory \cite{Wang:2000uj,Moore:2004tg} $\hat{q}$ is
 calculated
in the thermal field theory approach \cite{Weldon:1983jn} using the notation and
convention of \cite{LeBellac:1996}, generalizing the expression given by
Eq.~(\ref{eq:basic})
to include the finite temperature dependence. The coefficient $\hat{q}$ is expressed by
the thermal quark self-energy, however weighted by $q_\perp^2$,

\beq
\hat{q} = \frac{1}{4 p^0} {\rm Tr}~[ {p\slashh} \Sigma_w (p) ]~,
\label{sigma0}
\eeq
where explicitly

\beq
\Sigma_w (p) = g^2 C_F \int~ \frac{d^4q}{(2\pi)^4}~ q_\perp^2
~[\gamma^\mu {S\slash}_F^{>}(p') \gamma^\nu \Delta^{>}_{\mu \nu} ] ~.
\label{sigma1}
\eeq
 $g$ is the QCD coupling, $C_F = (N_c^2 -1)/2 N_c$ is the colour factor of the
 quark coupled to the gluon.
The hard quark with momentum $p' = p + q$ in the medium remains on-shell at
 temperature $T=0$,

\beq 
{S\slash}^{>}_F (p') = 2 \pi {p'\slash} \delta_{+} (p'^2) =
 2 \pi {p'\slash} \theta (p'^0) \delta (p'^2) ~.
\eeq 

\noindent
The gluon  propagates with momentum $q^\mu$ through the thermal medium,

\beq
\Delta^{>}_{\mu \nu} = - 2 (1 + f(q^0))~{\rm Im} ~ \Delta_{\mu \nu}~,
\label{KMS}
\eeq
with the Bose-Einstein distribution $ f(q^0) = ( \exp(q^0/T) -1)^{-1}$.
$\Delta_{\mu \nu}$ denotes the gluon propagator, which is treated in
detail in Appendix A.

In the relativistic case and for small momentum transfer,
 $-q^2/2 =  p \cdot p' \simeq q_{\perp}^2/2$,
and with
$p \cdot q = p^0 (q^0 - \vec{v} \cdot \vec{q}) \simeq 0$, $\vec{v} = \vec{p}/p^0 =
(0,0,1)$,
one may approximate

\beq
\hat{q} = - \frac{g^2 C_F}{2 p^0}~{\rm Im}~ \int  \frac{d^4q}{(2\pi)^4}~
q_\perp^2 ~ 2 \pi \delta_{+}((p+q)^2) ( 1 + f(q^0))
{\rm Tr}~[{p\slashh} \gamma^\mu ({p\slashh} + {q\slashh}) \gamma^\nu] \Delta_{\mu \nu}~,
\eeq
by

\beq
\hat{q} = - g^2 C_F~{\rm Im}~ \int  \frac{d^4q}{(2\pi)^4}~
q_\perp^2 ~  \frac{2T}{q^0} 2 \pi \delta (q^0 - \vec{v} \cdot \vec{q}) 
~\frac{p\cdot  \Delta \cdot p}{(p^0)^2}~,
\label{finaleq}
\eeq
when $q^0 << T$ and $q^0/p^0 \rightarrow 0$.
Concerning kinematics: $\vec{v} \cdot \vec{q} = \vert \vec{q} \vert
\cos\theta_{pq} = q^0$,  $ q_\perp^2 =  \vert \vec{q} \vert^2
\sin^2 \theta_{pq} = \vert \vec{q} \vert^2 (1 - x^2)$ with $ x = q^0/ \vert \vec{q} \vert$.

As already pointed out in \cite{Romatschke:2006bb} this equation 
(\ref{finaleq}) tells us that all
the information about the medium,
either isotropic or anisotropic, is contained in the imaginary part of the
gluon propagator (Appendix A).

\section{Hard-Thermal-Loop self-energy 
in an anisotropic plasma}
\label{htl-sec}

In the hard-thermal-loop approximation \cite{Braaten:1989mz}
the retarded gauge-field self-energy is
given by~\cite{ThomaMrow}
\beq
\Pi^{\mu \nu}(q)= g^2 \int \frac{d^3 {\vec{p}}}{(2\pi)^3} \,
v^{\mu} \frac{\partial n({\vec{p}})}{\partial p^\beta}
 \left( g^{\nu \beta} -
\frac{v^{\nu} q^\beta}{q\cdot v + i \epsilon}\right) \; ,
\label{self} 
\eeq
where $v^{\mu} \equiv (1,{\vec{p}}/|{\vec{p}}|)$ is a light-like vector
describing the propagation of a plasma particle in space-time. The
self-energy is symmetric, $\Pi^{\mu\nu}(q)=\Pi^{\nu\mu}(q)$, and
transverse, $q_\mu\Pi^{\mu\nu}(q)=0$.

In order to determine $\Pi^{\mu \nu}$
 the phase space distribution function has to be specified.
The  following form is used,
\begin{equation}
n({\vec{p}}) = n_{\rm iso}\left(\sqrt{{\vec{p}}^2+\xi({\vec{p}}\cdot{\vec{
n}})^2} \right) ~.  \label{eq:f_aniso}
\end{equation}
Thus, $n({\vec{p}})$ is obtained from an isotropic distribution $n_{\rm
iso}(|\vec{p}|)$ by removing particles with a large momentum component
along $\vec{n}$ \cite{Romatschke:2003yc,Romatschke:2003ms,Romatschke:2004jh}.
In a more general approach  the distribution
$n_{\rm iso}$ need not necessarily be thermal, as assumed here.

The functions $\alpha, \beta, \gamma$ and $\delta$,
introduced in the Appendix A in Eq.~(\ref{decomp})
are obtained from Eq.~(\ref{self}). 
Explicit expressions maybe found in
 \cite{Romatschke:2003yc,Romatschke:2003ms,Romatschke:2006bb}.

 \def\l{\left}
 \def\r{\right}

\subsection{ isotropic case}

For reference the case for $\xi = 0$ is first considered,
for which the propagator is given by Eq.~(\ref{eq:isop}),
containing
the unscreened transverse mode and the Debye screened
longitudinal one.
Therefore   dynamical screening due to Landau damping
\cite{LeBellac:1996}, as e.g. done in \cite{Wang:2000uj}, 
has to be included.

To leading logarithm order (LL) in $\ln{\frac{T}{m_D}} \simeq \ln{\frac{1}{g}}$ 
- in the following the only limit of interest -
one may keep only terms
 linear in $x $
in the denominator of the propagator
in Eq.~(\ref{eq:isop}), and approximate

\beq
{\rm Im}~ \l[ \frac{\hat{p} \cdot A \cdot \hat{p}}{q^2 -\Pi_T}
+ \frac{\omega^2 ~ \hat{p} \cdot B \cdot \hat{p}/q^2}{q^2 - \tilde{\Pi}_L}  \r]~, 
\eeq
with $\omega = q^0, \hat{p}^\mu = p^\mu/p^0$, by

\beq
{\rm Im}~ \l[ \frac{(1 - x^2)}{q^2 - i {\rm Im} \Pi_T}
+ \frac{1}{\vec{q}^2 + m_D^2 -i {\rm Im} \Pi_L}  \r] \simeq
\eeq
\beq
{\rm Im}~ \l[ \frac{1}{- \vec{q}^2 + i \pi m_D^2 x/4}
+ \frac{1}{\vec{q}^2 + m_D^2  +i \pi m_D^2 x/2}  \r] ~,
\eeq
\noindent
using Eq.~(\ref{impart}).

Performing the integrations to LL order with  $q = \vert \vec{q} \vert \leq T$
gives finally for  a hard quark jet 

\beq
\hat{q}_{iso} = 
\frac{g^2 C_F m_D^2 T}{2 \pi} \ln{\frac{T}{m_D}} ~,
\label{qiso}
\eeq
which coincides with the kinetic result Eq.~(\ref{refkin}) given in Appendix B.
Dynamical screening indeed guaranties this finite answer.

\subsection{anisotropic case}

\def\qh{\hat{q}_y}

In the hot anisotropic system $(\xi \ne 0)$ the gluon
propagator of the form of Eq.~(\ref{eq:anisop})
is the relevant one, such that

\beq
\hat{p} \cdot \Delta \cdot \hat{p}=
\Delta_A ~\left[ 1 - x^2 - \frac{x^2 \qh^2}{1- \qh^2} \right] +
\Delta_G ~\left[x^2 (q^2 - \alpha -\gamma) + \frac{x^2 \qh^2}{1- \qh^2}
(\omega^2 - \beta) - 2 x^2 \qh \delta\right] ~,
\label{ pDp}
\eeq
with the transverse momentum component 
$\qh = \vec{q} \cdot \vec{n}/|\vec{q}| = q_y/|\vec{q}|$
into the direction of anisotropy.

First, the contribution of $\Delta_A$ to $\hat{q}$ of Eq.~(\ref{finaleq})
is considered, which even to LL accuracy shows the possible presence
of the plasma instability. The contribution reads

\beq
\hat{q}_A = - \frac{g^2 C_F T}{4 \pi^3}
\int d\Omega_q \frac{1 - x^2}{x} \left[ 1 - x^2 - \frac{x^2 \qh^2}{1- \qh^2} \right]
~I(x, \alpha) ~,
\label{qA}
\eeq
noting that $x = \vec{v} \cdot \vec{q}/|\vec{q}|$ does depend only on angles,
as does $\qh$, which are integrated by the angular integration in 
(\ref{qA}).

In performing the $|\vec{q}|= q-$integration

\beq
I(x, \alpha) = {\rm Im} \int_0^T~ q^3 dq ~\Delta_A (x,q, \alpha) =
- {\rm Im} \int_0^T~  dq~
\frac{q^3}{q^2 (1-x^2) + {\rm Re} \alpha + i  {\rm Im}\alpha} ~,
\label{Imint}
\eeq

\noindent
it has to be kept in mind that ${\rm Im} \alpha$ is non-vanishing
because of Landau damping,
however, ${\rm Im} \alpha \propto x$ for $x \rightarrow 0$,
 and that ${\rm{Re}} \alpha$ is negative for
$\xi > 0$ in a range of $x$ values, $|x| < 1$,
as e.g. shown in Fig.~1a of \cite{Romatschke:2003ms}.
Therefore, in this range, including the point $x = 0$,
 there is a pole present in the integrand of
Eq.~(\ref{Imint}), characteristic of the anisotropy.

Evaluation gives

\bqa
I(x, \alpha) &=&  \frac{{\rm{Im}} \alpha}{2 (1-x^2)^2} ~
\left\{
\frac{1}{2} \ln{\frac{[T^2(1-x^2) + {\rm{Re}} \alpha]^2 
+ ({\rm{Im}}\alpha)^2}{|\alpha|^2}} \right.\nonumber \\
&+& \left.\frac{{\rm{Re}}\alpha}{{\rm{Im}}\alpha} ~
\left[\arctan\frac{{\rm{Im}}\alpha}{T^2 (1-x^2) + {\rm{Re}}\alpha}
 -\arctan \frac{{\rm{Im}}\alpha}{{\rm{Re}}\alpha}
+ \pi \Theta(- {\rm{Re}}\alpha)\right]
 \right\} ~.
\eqa

In the LL approximation this gives
\beq
I(x, \alpha) \simeq  \frac{{\rm{Im}}\alpha}{ (1-x^2)^2} ~
\left[\frac{1}{2} \ln{\frac{T}{m_D}} + \frac{\pi}{2} \frac{{\rm{Re}}\alpha}{{\rm{Im}}\alpha}
\Theta(- {\rm{Re}}\alpha) \right] ~,
\label{Ising}
\eeq
where the second term has to be kept, because it reflects a singularity
for ${\rm{Im}}\alpha \propto x \rightarrow 0$ due to the anisotropy $\xi > 0$.

To be explicit, the small $\xi$ behaviour is considered,
where \cite{Romatschke:2003ms}
\bqa
\alpha = && \Pi_T(x) + \xi \left\{ \frac{x^2}{6}
(5 \qh^2 -1) m_D^2  - \frac{\qh^2}{3} m_D^2\right.
\nonumber \\
&& +\left.\frac{1}{2} \Pi_T(x) [3 \qh^2 - 1 -x^2( 5 \qh^2 - 1)]
\right\}~,
\eqa
which shows that for $x \rightarrow 0$
\beq
{\rm{Re}} \alpha \simeq  - \frac{1}{3} \xi \qh^2 m_D^2 ~,
\label{real}
\eeq
indeed negative for $\xi > 0$,
and 
\beq
{\rm{Im}} \alpha \simeq - \frac{\pi}{4}~ x (1 - x^2) m_D^2~
\left\{ 1 + \frac{\xi}{2} [3 \qh^2 - 1 -x^2( 5 \qh^2 - 1)]
\right\}~.
\label{IMa}
\eeq

Following \cite{Romatschke:2006bb} the contribution from
the first term in Eq.~(\ref{Ising}) to $\hat{q}_A$ is denoted as regular.
 After performing the angular 
integrations in Eq.~(\ref{qA})
it leads at LL order with $T >>m_D$ to

\beq
\hat{q}_A^{reg} = \frac{g^2 C_F m_D^2 T}{8 \pi} \ln \frac{T}{m_D}
~(1 + O(\xi^2))~,
\label{qhxi}
\eeq
with no contribution at first order in $\xi$.

Turning to the regular contribution due to $\Delta_G$ one first obtains
\bqa 
\hat{q}_G = && \frac{g^2 C_F T}{8 \pi^3}~ \int  d\Omega_q ~x (1-x^2)~
 \nonumber \\ && \times ~{\rm Im}
 \int  dq^2 q^2\frac{~ ( 1-x^2) q^2 + \alpha + \gamma  
- \frac{x^2 \qh^2}{1- \qh^2}(q^2 - \tilde{\beta})
+ 2 \qh \tilde{\delta} }
{x^2 (1-x^2)(\tilde{\beta} - q^2) q^2 + x^2 (\alpha + \gamma)(\tilde{\beta} -
  q^2) - \tilde{\delta}^2 ( 1 - \qh^2)} ~.
\label{long}
\eqa
The regular LL part reads after the $q-$integration
\beq
 \hat{q}_G^{reg} = - \frac{g^2 C_F T}{4 \pi^3} \ln\frac{T}{m_D}~ \int 
 d\Omega_q ~ \frac{1}{x (1-x^2)}~
[ (1-x^2) {\rm Im} \tilde{\beta} + 2 (1-x^2) \qh {\rm Im}\tilde{\delta}
+ \frac{x^2 \qh^2}{1- \qh^2} {\rm Im}(\alpha + \gamma) ] ~,
\label{long2}
\eeq
with (\ref{IMa}) and 
 
\bqa
{\rm Im} \tilde{\beta} &=& - \frac{\pi}{2} x m_D^2~ 
{\lbrace} 1 + \xi~ [ 2 \qh^2 - 1 - x^2 (3 \qh^2 -1) ]
{\rbrace} ~, \nonumber \\ 
{\rm Im}(\alpha + \gamma) &=&  - \frac{\pi}{4} x (1- x^2) m_D^2~   
{\lbrace} 1 + \frac{\xi}{2} [5 \qh^2 - 3 - x^2 (7 \qh^2 -3) ] {\rbrace}
 ~,  \nonumber \\
{\rm Im}\tilde{\delta} &=& - {\frac{\pi}{4}} x (1- x^2) m_D^2~
 \xi~ ( 1 - 4 x^2 ) \qh   ~,
\label{ims}
\eqa
taken from \cite{Romatschke:2003ms,Romatschke:2006bb}.

The angular integration in Eq.~(\ref{long2}) then gives

\beq
\hat{q}_G^{reg} = \frac{3 g^2 C_F m_D^2 T}{8 \pi}
~\ln \frac{T}{m_D}~(1 + O(\xi^2)).
\label{qhg}
\eeq

Summing the two terms Eqs.~(\ref{qhxi}) and (\ref{qhg})
the LL transport coefficient in the limit of small $\xi$ up to $O(\xi)$
becomes
\beq
\hat{q}_{anisio}^{reg} = \hat{q}_A^{reg} +  \hat{q}_G^{reg} =
\frac{g^2 C_F m_D^2 T}{2 \pi}~ \ln\frac{T}{m_D}~ (1 + O(\xi^2)).
\label{regqhat}
\eeq

For $\xi =0$ the value for $\hat{q}_{iso}$, Eq.~(\ref{qiso}),
is recovered.
 This result  for $\hat{q}_{ansio}^{reg}$ in Eq.~(\ref{regqhat}) does  agree
 with the result given for
$\kappa_{\perp}^{reg} + \kappa_z^{reg}$ in Eq.~(14) of \cite{Romatschke:2006bb}
when the velocity $v$ is taken to be $v=1$ for a massless quark jet.
Because of this agreement, which is not unexpected, one may try to extrapolate
$\hat{q}_{aniso}^{reg}$ for $v \rightarrow 1$ from the numerical values
of $\kappa_{\perp,z}^{reg}$ summarized in Table~1 of \cite{Romatschke:2006bb} 
even for large values of $\xi$ by
\beq
\frac{\hat{q}_{aniso}^{reg}}{\hat{q}_{iso}}
\simeq 1 ~,
\label{extra}
\eeq
almost independent of $\xi \le 100$.

Next the anomalous contribution \cite{Romatschke:2006bb} due to the second
term of Eq.~(\ref{Ising}) is evaluated. In LL order only the behaviour
for $x \rightarrow 0$ is relevant. With Eq.~(\ref{real}) it gives
\beq
\hat{q}_A^{anom} \simeq  \frac{g^2 C_F m_D^2 T}{24 \pi^2}
\xi~ \int d\Omega_q \frac{\qh^2}{x} ~,
\label{anomq}
\eeq 
inducing a logarithmic singularity with
 $x = \cos \theta_{pq}$.
The contribution to $\hat{q}_G^{anom}$ starts at $O(\xi^2)$.

\noindent
A short way to see the anomalous contribution is to sum the singular parts in
the gluon propagator for the static case in terms of the masses,
here explicitly quoted from \cite{Romatschke:2003ms}
in the limit of small anisotropy $\xi \rightarrow 0$ scaled with respect to
the Debye mass Eq.~(\ref{Dmass}), i.e. $ \hat{m} = m/m_D$,
\bqa
\hat m_\alpha^2 &=& - {\xi\over6}(1 + \cos 2\theta_n) \; , \nonumber \\
\hat m_\gamma^2 &=& {\xi\over3}\sin^2\theta_n \; , 
\label{delta}
\eqa
and

\bqa\label{masses}
\hat m_+^2 &=& 1 + {\xi\over6}(3 \cos 2\theta_n-1) \; , \nonumber \\
\hat m_-^2 &=& -{\xi\over3}\cos2\theta_n \; ,
\eqa
where the angle is given by $\cos\theta_n = \qh$.

Since $m^2_{\alpha}$ and $m_{-}^2$ are negative the pole contributions
(with the Feynman prescription) read
\beq
\hat{q}^{anom} = \frac{ g^2 C_F  T}{8 \pi^3}
~{\rm Im} \int d\Omega_q
\int\frac{q^2 dq^2}{x} 
\left[ \frac{1}{q^2 + m_\alpha^2 -i \epsilon}
- \frac{1}{m_+^2 - m_{-}^2} 
\frac{m_{-}^2 - m_\alpha^2 - m_\gamma^2}{q^2 + m_{-}^2 -i \epsilon} 
\right] ~.
\label{massanom}
\eeq
The term of $O(\xi)$ comes from the first pole term 
and leads to Eq.~(\ref{anomq}), whereas the second term is obviously
of $O(\xi^2)$.

The question about the treatment  of the  logarithmic singularity in
Eq.~(\ref{anomq})  arises. At least three possibilities to cut-off the
singularity  may be discussed:

(i) One may 
follow the detailed and plausible 
arguments given in \cite{Romatschke:2006bb}
that this soft singularity is screened by
$O(g^3)$ terms in the gluon propagator, i.e. beyond the HTL approximation
under discussion. It leads to the replacement of  ${\rm{Im}} \alpha$
in the  second term in the denominator of Eq.~(\ref{Ising}) by
${\rm{Im}} \alpha  \sim x \rightarrow x + c g$, i.e. it is suggestive to cut the
singularity in (\ref{anomq}) by
\beq
\xi \int \frac{d x}{x} \rightarrow 2 \xi \int_0 \frac{d x}{x + cg}
\sim 2 \xi \ln\frac{1}{g} \sim 2 \xi \ln \frac{T}{m_D} ~.
\label{save}
\eeq
This way a finite result is obtained,
\beq
\hat{q}_A^{anom} \simeq  \frac{g^2 C_F m_D^2 T}{2 \pi}
\ln \frac{T}{m_D} ~ \frac{\xi}{6} ~,
\label{anomf}
\eeq 
which  shows a positive, but weak dependence on $\xi$ as a sign of the
anisotropy for $\xi > 0$..

(ii) The origin of the $1/x$ singularity is traced back to the Bose-Einstein
distribution $f(q^0) \sim T/q^0$ in Eq.~(\ref{KMS}). Pragmatically,
in the anisotropic case, this behaviour could be modified by
$q^0 \rightarrow \sqrt{(q^0)^2 + \xi (\vec{n} \cdot \vec{q})^2}
= q~\sqrt{x^2 + \xi \qh^2}, ~ q^0 > 0$. On mass-shell this replacement gives
the distribution in  Eq.~(\ref{eq:f_aniso}),  and  leads  to
\beq
\xi \int \frac{d x}{x} \rightarrow 2 \xi \ln \frac{1}{\xi} ~.
\label{save2}
\eeq

(iii) To form the anisotropic configuration in momentum space 
a characteristic time scale is present of the order
$\tau_c \sim O(1/g \xi T)$, for not to large $\xi$.
It is then natural to cut the energies of the constituents in the heat bath
by $|q^0| \ge 1/\tau_c > g \xi T$, and
\beq
\xi \int \frac{d x}{x} \rightarrow 2 \xi \ln \frac{1}{g \xi} ~.
\label{save3}
\eeq

In summary all three options Eqs.~(\ref{save} - \ref{save3}) lead to a positive
contribution of $O(\xi)$ at LL order to  $\hat{q}_{aniso}$.

\section {Conclusion}

Because of the approximations, keeping only leading logarithmic order terms, 
a detailed quantitative study is not aimed.
Qualitatively, however, the  main result obtained strongly indicates that
the instability due to the anisotropy leads to
$\hat{q}_{aniso} > \hat{q}_{iso}$.
For small anisotropies it turns out to be an  effect of $O(\xi)$,
which should   be taken into account
in the future phenomenological comparison with experimental jet quenching data,
although the actual numerical value depends on the full treatment including the
non-leading terms.

\vspace{0.3cm}

Independent information on $\hat{q}_{aniso}$ maybe found in the following
references:

\noindent
i) A result $\hat{q}_{aniso} \ge \hat{q}_{iso}$ is found in a numerical 
simulation in \cite{Dumitru:2007rp} treating jet broadening in an unstable
non-Abelian ($SU(2)$) plasma. At early evolution times 
$\hat{q}_{aniso} \simeq \hat{q}_{iso} \simeq 2.2 ~GeV^2/fm$, whereas at times
when the instability is acting there are numerical indications that 
$\hat{q}_{aniso}$
increases when compared to $\hat{q}_{iso}$.

\noindent
ii) The transport coefficient $\hat{q}$ of a fast parton (gluon) propagating
in an expanding turbulent quark-gluon plasma is estimated in
\cite{Asakawa:2006tc,Asakawa:2006jn,Muller:2007rs}. The
 coefficient  called anomalous in the quoted references 
(but not to be confused with the one in Eq.~(\ref{qA})) is
parametrically given by

\beq
\hat{q}_A \simeq g \xi^{\frac{3}{2}} m_D^2 T ~,
\label{Muller}
\eeq
therefore $ \hat{q}_A / \hat{q}_{iso} \simeq \xi^{\frac{3}{2}}/g > 1$,
even significantly larger than one for small coupling $g$ (assuming $\xi$
not very small).

\vspace{0.2cm}
These results together indeed show that $\hat{q}_{aniso} > \hat{q}_{iso}$.

\vspace{0.3cm}

\noindent
In $A-A$ collisions the parton saturation scale $Q_s$ 
\cite{Mueller:2001fv} is for a large nucleus $A$
related to the jet quenching parameter (of gluons)  $\hat{q}$
by $Q^2_s = 2 \hat{q} \sqrt{R^2- b^2}$ \cite{Mueller:1999wm,Mueller:2002kw,
Baier:2002tc,Solana:2007sw}, where $R$ is the radius of the nucleus  and $b$ is the impact
parameter of the collision. Because $\hat{q}$ is affected by anisotropies in
$A-A$ collisions, it is not obvious that
$Q_s$ determined from small$-x$ processes in proton/deuteron - 
nucleus collisions is indeed the same as in $A-A$ collisions.

\acknowledgments

R.~B. is  indepted to Paul Romatschke for 
usefull suggestions and crucial clarifications.
We thank Michael Strickland for comments especially on the results given in
\cite{Dumitru:2007rp}.

\section{Appendix}

\subsection{Propagator in covariant gauge in an anisotropic plasma}

Here  we summarize the results for the gluon self-energy $\Pi^{\mu\nu}$
and the gluon propagator $i\Delta^{\mu\nu}_{ab}$ (diagonal in colour)
in covariant gauge in an anisotropic medium.

The self-energy is decomposed as
\begin{equation}\label{decomp}
\Pi^{\mu\nu}=\alpha A^{\mu\nu}+\beta B^{\mu\nu} + \gamma
C^{\mu\nu} + \delta D^{\mu\nu} ~.
\end{equation}

The tensor basis of $\Pi^{\mu\nu}$ is
determined explicitly for anisotropic systems in
~\cite{Romatschke:2003ms}. We use here a  basis 
 appropriate for the  covariant gauge, and define
\beq \label{eq:A_munu}
 A^{\mu \nu}= -P^{\mu\nu} +\frac{\tilde{K}^\mu \tilde{K}^\nu}{\tilde{K}^2} =
 -g^{\mu\nu}+\frac{q^\mu
q^\nu}{q^2}+\frac{\tilde{K}^\mu \tilde{K}^\nu}{\tilde{K}^2}
\eeq
\beq\label{eq:B_munu}
B^{\mu \nu}= -\frac{q^2}{(q\cdot u)^2}~
 \frac{\tilde{K}^\mu \tilde{K}^\nu}{\tilde{K}^2}
\eeq
\beq\label{eq:C_munu}
C^{\mu \nu}= -
\frac{\tilde{n}^\mu\tilde{n}^\nu}{\tilde{n}^2}
\eeq
\beq   \label{eq:D_munu}
D^{\mu \nu}= 
\tilde{K}^\mu \tilde{n}^\nu + \tilde{K}^\nu \tilde{n}^\mu ~,
\eeq
where $u^{\mu}$ is the heat-bath vector, which in the local rest frame
is given by $u^{\mu}=(1,0,0,0)$, and
\begin{equation}\label{eq:Kmu}
\tilde{K}^\mu=q^{\mu}-\frac{q^2}{(q \cdot u)} \,u^\mu ~.
\end{equation}

The direction of anisotropy in momentum space is given by the
vector
\begin{equation}
n^{\mu}=(0,{\vec{n}})~.
\end{equation}
%
Defining the vector
\begin{equation}
\tilde{q}^\mu = q^\mu - (q \cdot u) u^\mu ~,
\end{equation}
and the projector
\begin{equation}
\tilde{P}^{\mu \nu} = g^{\mu\nu} - u^\mu u^\nu 
- \frac{\tilde{q}^\mu \tilde{q}^\nu}{\tilde{q}^2}~,
\end{equation}
introduces the vector
\begin{equation}
\tilde{n}^\mu = \tilde{P}^{\mu\nu} n_\nu~.
\end{equation}
Both vectors $\tilde{n}^\mu$ and $\tilde{K}^\mu$
 are  orthogonal to $q^\mu$, and $\tilde{K} \cdot \tilde{n} = 0$.

The inverse propagator in covariant gauge is then
\begin{equation}
\left(\Delta^{-1}\right)^{\mu \nu}(q,\xi)= -q^2 g^{\mu \nu} +q^\mu q^\nu
-\Pi^{\mu\nu}(q,\xi)-\frac{1}{\lambda}q^\mu q^\nu ~,
\end{equation}
where $\lambda$ is the gauge fixing parameter. Upon inversion 
\cite{Kobes:1990dc}, the
propagator is finally  written as, with $\omega = (q \cdot u)$,
\begin{equation}\label{eq:anisop}
\Delta^{\mu\nu} = \Delta_{A} \left[A^{\mu\nu} - C^{\mu\nu}\right] + 
\Delta_{G}\left[(q^2-\alpha-\gamma)\frac{\omega^4}{q^4}B^{\mu\nu} +
(\omega^2-\beta)C^{\mu\nu} + \delta\frac{\omega^2}{q^2}D^{\mu\nu}\right] - 
\frac{\lambda}{q^4}q^\mu q^\nu ~,
\end{equation}
where
\begin{equation}
\Delta^{-1}_{A} = (q^2-\alpha) ~,
\label{deltaa}
\end{equation}
and
\begin{equation}
\Delta^{-1}_{G} = (q^2-\alpha-\gamma)(\omega^2-\beta) + 
\delta^2 \vec{q}~^2 \tilde{n}^2~.
\label{deltag}
\end{equation}
For $\xi=0$, we recover the isotropic propagator in covariant gauge
\cite{Braaten:1989mz}
\begin{equation}\label{eq:isop}
\Delta^{\mu\nu}_{iso} = \frac{1}{q^2-\Pi_{T}}A^{\mu\nu} + 
\frac{1}{(q^2-\tilde{\Pi}_L)}\frac{\omega^2}{q^2}B^{\mu\nu} - 
\frac{\lambda}{q^4}q^\mu q^\nu ~,
\end{equation}
where $\tilde{\Pi}_L = \frac{q^2}{\vec{q}{^2}} \Pi_L$ and \cite{LeBellac:1996}
\bqa
\Pi_T(\omega,\vec{ q})&=&\frac{m_D^2}{2}
\frac{\omega^2}{\vec{ q}^2}\left[1-
\frac{\omega^2-{\vec{q}}^2}{2 \omega |\vec{ q}|} \log{\frac{\omega+|\vec{
q}|}{\omega-|\vec{ q}|}}\right] ~, \nonumber\\
\Pi_L(\omega,\vec{ q})&=&m_D^2\left[\frac{\omega}{2 |\vec{q}|}
\log{\frac{\omega+|\vec{ q}|}{\omega-|\vec{q}|}}-1\right]~.
\label{isoprop}
\eqa

\noindent
In the soft $\omega$ limit we remind that

\beq\label{impart} 
\Pi_T \rightarrow -i \pi \frac{\omega}{4 \vert \vec{ q} \vert}
(1 - \frac{\omega^2}{\vert \vec{q} \vert^2}) m_D^2~,
~~\Pi_L \rightarrow -m_D^2 - i 
\pi\frac{\omega}{2 \vert \vec{q} \vert} m_D^2  ~.
\eeq

\noindent
 $m_D$ is the isotropic Debye mass,
which for a thermal system at temperature $T$ 
with $N_c$ colours and $N_f$ light quark flavours 
 is given by
\beq\label{Dmass}
 m_D^2=\frac{g^2 T^2}{3}\left(N_c +\frac{N_f}{2}\right).
\eeq

Eq.~(\ref{eq:anisop}) agrees with the expression given
in \cite{Dumitru:2007hy}, when the following relations are used,

\begin{equation}
\tilde{K}^\mu = - \frac{q^2}{(q \cdot u)} \tilde{m}^\mu ~,
\end{equation}
and
\begin{equation}
\tilde{n}^\mu = \tilde{n}_{DGS}^\mu
-\frac{(q \cdot n)}{\tilde{K}^2} \tilde{K}^\mu~, 
\end{equation} 
where the tilde vector in \cite{Dumitru:2007hy}
is given by
\begin{equation}
\tilde{n}_{DGS}^\mu = n^\mu - \frac{q \cdot n}{q^2} q^\mu~.
\end{equation}

\subsection{Kinetic theory approach}

\def\M{\bar{\cal M}}

In the case of an isotropic medium the transport coefficient $\hat{q}$
maybe calculated using Eq.~(\ref{eq:basic}), where the
differential cross section is related to the spin and colour summed (averaged)
squared matrix element by

\beq\label{isocross}
\frac{d \sigma}{d^2 q_\perp} = \frac{1}{16 \pi^2 s^2} {\vert {\M} \vert}^2~,
\eeq 
 with $s$  the invariant energy squared.

When properly  taking quantum statistics (Pauli-blocking and Bose-Einstein
 enhancement)
 into account,
it is appropriate to modify (\ref{eq:basic}) as
\beq
\rho{\vert {\M} \vert}^2  \rightarrow \int~\frac{d^3k}{(2 \pi)^3}
\l[ 4N_c N_f  n_{FD} (1 - n_{FD}){\vert {\M} \vert}^2_{qq'}
+ 2 (N_c^2 -1) n_{BE} (1 + n_{BE}){\vert {\M} \vert}^2_{qG} \r] ~,
\eeq
together with the scattering of the quark jet with the thermalised quarks
and gluons in the hot medium, distributed according to the functions $n_{FD}$
and $n_{BE}$, respectively.

At leading order,  after screening 
the effective forward scattering amplitudes \cite{Heiselberg} 
by the Debye mass
 $1/q_\perp^2 \rightarrow 1/(q_\perp^2 + m_D^2)$,
the result is
\beq
\hat{q}\vert_{kinetic} = \frac{g^2  C_F m_D^2 T}{2\pi} \ln{\frac{T}{m_D}} ~,
\label{refkin}
\eeq
in agreement with Eq.~(\ref{qiso}). This result agrees with the LL one
  for 
$d<p^2_{\perp}>/dt = 2~ \kappa_T$ calculated in
\cite{Moore:2004tg} (Eq.~B32) taking the quark velocity to be $v =1$.

\end{document}